\documentclass[manuscript]{acmart}
\AtBeginDocument{%
  }

\setcopyright{acmlicensed}
\copyrightyear{2025}
\acmYear{2025}
\acmDOI{XXXXXXX.XXXXXXX}
\acmISBN{978-1-4503-XXXX-X/2018/06}

\usepackage{tikz}
\usepackage{threeparttable}
\newcommand{\bcirc}[1]{%
  \tikz[baseline=(char.base)]{
    \node[shape=circle, fill=black, text=white, inner sep=0.3pt, minimum size=1.6ex] (char) {%
      \fontsize{0.8em}{0.8em}\selectfont #1};
  }%
}
\usepackage{multirow} 

\begin{document}

\title{Hemlet: A Heterogeneous Compute-in-Memory Architecture for Vision Transformers with Group-Level Parallelism}

\author{Cong Wang}
\authornote{Both authors contributed equally to this research.}
\email{cwang841@connect.hkust-gz.edu.cn}

\author{Zexin Fu}
\authornotemark[1]
\email{zexin.fu@connect.hkust-gz.edu.cn}
\affiliation{%
  \institution{The Hong Kong University of Science and Technology (Guangzhou)}
  \country{China}
}

\author{Jiayi Huang}
\email{hjy@hkust-gz.edu.cn}
\affiliation{%
  \institution{The Hong Kong University of Science and Technology (Guangzhou)}
  \country{China}
}

\author{Shanshi Huang}
\email{shanshihuang@hkust-gz.edu.cn}
\affiliation{%
  \institution{The Hong Kong University of Science and Technology (Guangzhou)}
  \country{China}
}


\begin{abstract}
 Vision Transformers (ViTs) have established new performance benchmarks in vision tasks such as image recognition and object detection. However, these advancements come with significant demands for memory and computational resources, presenting challenges for hardware deployment. Heterogeneous compute-in-memory (CIM) accelerators have emerged as a promising solution for enabling energy-efficient deployment of ViTs. Despite this potential, monolithic CIM-based designs face scalability issues due to the size limitations of a single chip. To address this challenge, emerging chiplet-based techniques offer a more scalable alternative. However, chiplet designs come with their own costs, as they introduce expensive communication, which can hinder improvements in throughput.

This work introduces Hemlet—a heterogeneous CIM chiplet system designed to accelerate ViT. Hemlet facilitates flexible resource scaling through the integration of heterogeneous analog CIM (ACIM), digital CIM (DCIM), and Intermediate Data Process (IDP) chiplets. To improve throughput while reducing communication overhead, it employs a group-level parallelism (GLP) mapping strategy and system-level dataflow optimization, achieving speedups ranging from 2.41× to 5.74× across various hardware configurations within the chiplet system. Our evaluation results demonstrate that Hemlet can achieve a throughput of 9.56 TOPS with an energy efficiency of 4.98 TOPS/W.
\end{abstract}







\maketitle

\section{Introduction}
The Vision Transformer (ViT) \cite{vit} has emerged as a foundational model that replaces convolutional neural network (CNN) with Transformer blocks, achieving state-of-the-art performance in computer vision. However, this improvement in performance comes with significantly higher computational demands, primarily driven by massive vector-matrix multiplications (VMMs). In addition to its computational intensity, ViT models are characterized by larger parameter sizes and require substantial frequent memory access to weights during inference. This frequent data transfer limits both throughput and energy efficiency on traditional von Neumann architectures, presenting a challenge commonly referred to as the "memory wall" bottleneck.

The compute-in-memory (CIM) paradigm \cite{chi2016prime,shafiee2016isaac,AEPE,yu2018neuro} addresses this bottleneck by performing VMMs directly within memory arrays, thereby reducing data movement and enhancing overall efficiency. Under the CIM paradigm, heterogeneous architectures that combine emerging non-volatile memory (eNVM)-based and SRAM-based CIM designs have been widely explored to address the diverse computation and storage demands of Transformer workloads\cite{x-former, liu2023hardsea,laguna2022hardwaresoftware,jain2023heterogeneous}. This heterogeneous design is primarily driven by the intrinsic trade-offs between different memory technologies. On one hand, eNVM devices, such as resistive random access memory (RRAM), offer high storage density and high energy efficiency \cite{shimeng2021review,jiang2023computememory}. However, their limited write endurance and high write overheads pose challenges for handling dynamically generated VMMs within multi-head attention (MHA)\cite{chakraborty2020resistive,zahoor2020resistive,ankit2020panther}. On the other hand, SRAM-based CIM designs provide faster access and better write endurance, but suffer from lower storage density and higher area overhead. To leverage the complementary strengths of these memory technologies, heterogeneous CIM systems typically employ eNVM-based CIM for static VMMs, which utilize fixed pretrained weights. Meanwhile, SRAM-based CIM is used to handle dynamic VMMs that require frequent operand updates during inference. This division of roles allows for efficient support of both static and runtime-generated computations in Transformer inference.

While the heterogeneous design can effectively address the demands of Transformers and maximize performance by statically storing all pretrained weights on-chip, the large parameter sizes of these models present significant challenges. Specifically, storing all pretrained weights in eNVM-based arrays substantially increases chip area, leading to considerable issues related to fabrication cost and yield \cite{chipletcost}. Furthermore, these designs are often optimized for specific workloads (e.g., ViTs, or customized Transformer variants). As model architectures evolve, these tightly coupled designs necessitate complete redesign, verification, and fabrication cycles, resulting in high recurring engineering costs and extended development timelines.
To address the challenges posed by monolithic heterogeneous CIM architectures, chiplet-based design emerges as a promising solution. By integrating multiple chiplets within a single package, these systems enhance both modularity and scalability \cite{shao2019simba,caichipletmapping}. Recent studies have started to investigate the potential of CIM chiplet platforms for accelerating DNN workloads \cite{siam,biglittle,sharma2023heterogeneous,dhingra2025atleusacceleratingtransformersedge,cimchiplet}.


However, the scalability offered by chiplet-based systems comes at a cost. Chiplets connected via network-on-package (NoP) face greater data transmission overhead compared to network-on-chip (NoC) on a single chip\cite{LVDS,XSRSerDes,caichipletmapping}. Therefore, optimizing the dataflow to reduce transmission overhead is crucial to maintaining performance in chiplet-based systems. Moreover, we observed that the conventional layer-wise mapping method used for ACIM fails to fully utilize the available hardware resources in real implementations, thereby limiting throughput. This presents an opportunity to compensate for the performance degradation caused by chiplet partitioning.




In this work, we introduce Hemlet—a heterogeneous CIM chiplet-based system designed for efficient inference of ViTs. Hemlet integrates a group-level parallelism (GLP) weight mapping strategy with optimized inter- and intra-chiplet dataflow. Our key contributions are summarized as follows:

\begin{enumerate}
    \item We propose \textbf{Hemlet}, a heterogeneous CIM chiplet-based architecture, integrating three types of chiplets: Analog CIM (ACIM), Digital CIM (DCIM), and Intermediate Data Process (IDP). By decoupling functionality across chiplets while maintaining intra-chiplet uniformity, Hemlet supports both architectural scalability and manufacturing efficiency.


    
    \item  A novel group-level parallelism (GLP) mapping method is proposed. In this approach, ACIM columns that share the same ADC are defined as a "Group" and parallelism among them is exploited to fully utilize the hardware resources. To facilitate efficient mapping of ViT models to Groups, we introduce an intermediate representation called the \textit{GLP\_LayerSet}, upon which we have developed a four-stage mapping procedure.

    \item System-level dataflow is optimized to minimize inter-chiplet communication. A fine-grained pipeline is utilized for the communication and computation of static VMMs, effectively masking inter-chiplet communication between ACIM and IDP. Furthermore, the DCIM chiplet architecture is co-optimized with blocked dynamic VMMs, eliminating global buffer access across chiplets. These designs significantly enhance throughput in Hemlet’s heterogeneous chiplet architecture.
    \item We conduct comprehensive evaluations of Hemlet across a range of ViT workloads and system configurations. The results show that {GLP} significantly improves ACIM throughput by maximizing hardware utilization. Additionally, the proposed {Hemlet} architecture, which combines GLP with system-level dataflow optimization, consistently achieves speedups across diverse hardware settings. Moreover, Hemlet demonstrates competitive performance compared to state-of-the-art accelerators while offering enhanced flexibility and scalability.
\end{enumerate}

The remainder of this paper is organized as follows: Section \ref{sec:background} introduces the preliminaries of compute-in-memory and chiplet-based architectures. Section \ref{sec:architecture} presents the proposed Hemlet architecture, followed by the GLP weight mapping strategy in Section \ref{sec:glp-big-section}. Section \ref{sec:dataflow} details the system-level dataflow optimization. Some evaluation results are presented in Section \ref{sec:evaluation} and a conclusion is drawn in Section \ref{sec:conclusion}.

\section{Preliminary}
\label{sec:background}
This section introduces background on CIM technology, including ACIM, DCIM, and their heterogeneous integration for Transformer acceleration, as well as the basics of chiplet-based architecture.

\subsection{Compute-in-Memory}

\begin{figure}
    \centering
\includegraphics[width=3.2in]{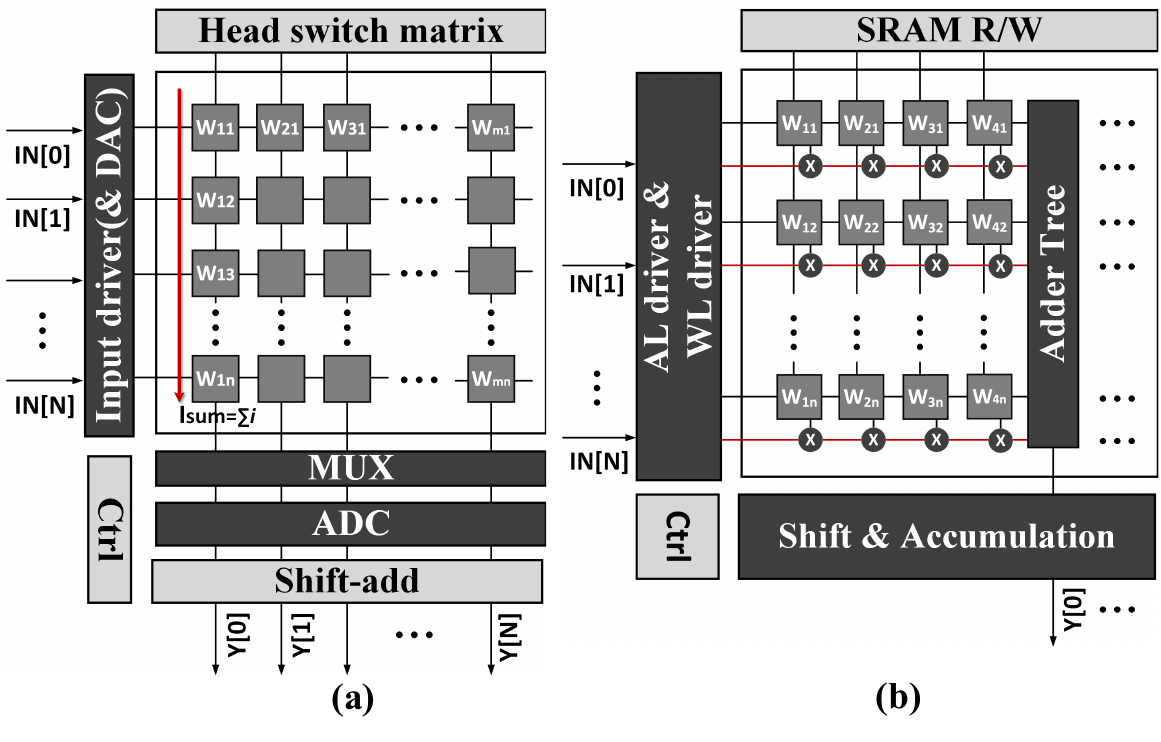}
    \caption{Architecture of (a) Analog CIM  and (b) Digital CIM }
    \label{fig:cimarch}
\end{figure}

\textbf{Analog Compute-in-Memory (ACIM):} 
Conductance-based ACIM macros perform VMMs by leveraging the resistive properties of emerging non-volatile memory (eNVM) devices such as resistive random access memory (RRAM)\cite{chen2019cmos,yao2020fully}, phase change memory (PCM)\cite{ambrogio2023analog,kim2019PCM,le202364}, and ferroelectric field-effect transistors (FeFET)\cite{yin2018ferroelectric,ni2018fefet}. This approach is highly appealing for energy-efficient computing due to its high-precision analog multiply-and-accumulate (MAC) capabilities and substantial storage density. As illustrated in Fig.\ref{fig:cimarch} (a), matrix elements are encoded into the conductance states of the eNVM cells. Depending on the specific characteristics of the memory devices, a single eNVM cell can represent multi-bit weights across multiple conductance levels, accommodating from 1 to 7 bits per cell \cite{ankit2019puma}.  Input vectors are converted into different voltages levels and applied to the arrays via wordlines (WLs), with the resulting column currents representing the MAC outputs according to Kirchhoff’s Law. In general, these inputs can also be high precision, depending on the number of voltage levels employed\cite{zhang2022pimca,lee202328,zhang2023macc}. The resulting analog currents are typically converted back to the digital domain through analog-to-digital converters (ADCs) located at the end of the bitlines for further processing. ACIM has been widely used in neural network inference accelerators\cite{chi2016prime,AEPE,shafiee2016isaac,retransformer,ReBert,lu2023rram}, as the VMMs dominate in these applications and pre-trained weights can be programmed into the ACIM array prior to inference, remaining unchanged throughout the process. While ACIM is well-suited for weight stationary implementations, it is not ideal for applications requiring frequent weight updates due to high write costs and endurance limitations.

\textbf{Digital Compute-in-Memory (DCIM):}
DCIM, built on SRAM-based memory arrays and featuring in-place digital computation, is evolving rapidly due to its flexibility and scalability to advanced technology nodes\cite{fengbin23isscc,dcim25jssc,autodcim,dac24dcim}. Fig. \ref{fig:cimarch} (b) illustrates the basic architecture of a DCIM array. These arrays consist of DCIM columns that share wordlines (WL) and activation lines (AL). Within each DCIM column, n-bit weights stored in SRAM cells are multiplied with 1-bit inputs on AL using local multipliers for each cycle. The resulting $1\times n$-bit products are accumulated along the columns through a digital adder tree. To support multi-bit input processing, outputs from different cycles are accumulated using shift \& accumulation (S/A). Compared to ACIM, DCIM offers much lower rewrite overhead due to the SRAM-based arrays and achieves higher throughput by eliminating the need for ADC conversion. However, it exhibits lower storage density and less energy efficiency.

\begin{figure}
    \centering
    \includegraphics[width=3.2in] {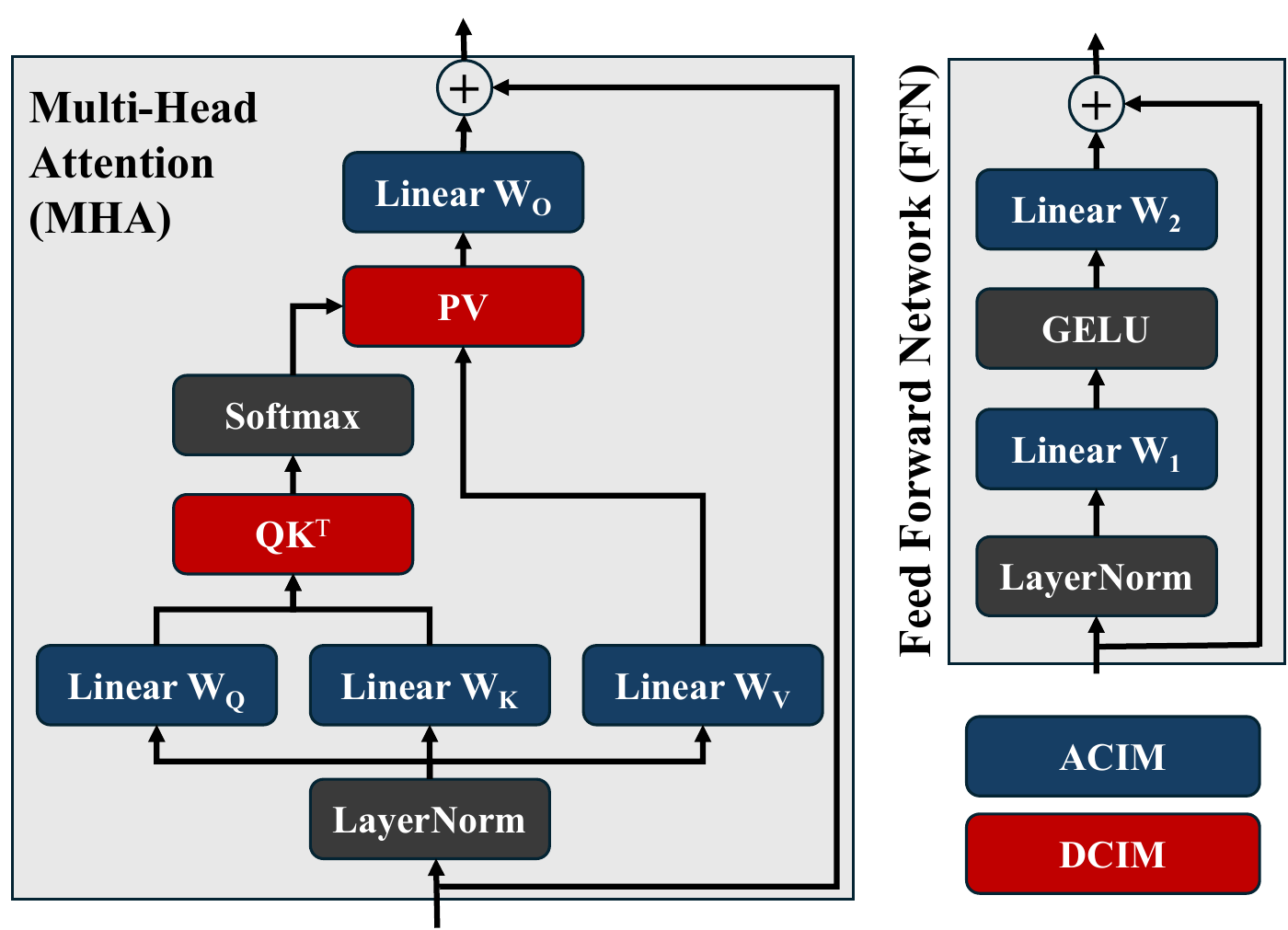}
    \caption{Heterogeneous CIM for Vision Transformer Computation}
    \label{fig:transformer}
\end{figure}

\textbf{Heterogeneous CIM for Transformers:}
ACIM is particularly effective for CNNs because the weights used in VMMs are pre-trained, thus eliminating the need for overwriting. However, the scenario shifts with Transformer-based networks, where certain VMMs related to attention are generated on-the-fly, necessitating rewrites in ACIM-only implementations. On the other side, the rapid scaling of operations in Transformers demands high energy efficiency, making DCIM-only accelerators less optimal. Consequently, efficient Transformer inference calls for a heterogeneous integration of ACIM (for static VMMs) and DCIM (for dynamic VMMs)~\cite{liu2023hardsea,li2023h3datten,fu2025optimizing}. Fig.~\ref{fig:transformer} illustrates how computations within a Transformer block are mapped onto the heterogeneous CIM system. Specifically, blue blocks denote the workload executed on ACIM, including the linear projections $\mathbf{W_Q}$, $\mathbf{W_K}$, $\mathbf{W_V}$, and $\mathbf{W_O}$ in multi-head attention (MHA), as well as the two fully connected layers $\mathbf{W_1}$ and $\mathbf{W_2}$ in the feed-forward network (FFN). Red blocks correspond to dynamic VMMs, namely $\mathbf{Q}\mathbf{K}^T$ and $\mathbf{P}\mathbf{V}$ in MHA, which are handled by DCIM.

\label{sec:acim}

\subsection{Chiplet architecture}

Chiplet architecture is an advanced packaging solution, which integrates multiple dies on an organic substrate or a silicon interposer within a single package\cite{zeppelin,siinterposer,amdchiplet}. Rather than fabricating one large monolithic chip, the system is partitioned into smaller, self‑contained “chiplets” that are interconnected via a network-on-package (NoP). By partitioning a large monolithic die into smaller chiplets, overall manufacturing yield increases dramatically, and the advanced substrates allow total package areas beyond the lithographic reticle limit, enabling larger systems than a single monolithic die. These characteristics directly overcome the size limitations inherent to monolithic chips and align seamlessly with the hardware resource requirements of large‑scale deep neural network (DNN) workloads\cite{shao2019simba,caichipletmapping,biglittle,cimchiplet}.

In addition to overcoming monolithic die size limits, chiplet-based designs offer inherent \textbf{scalability}: system capacity can be tuned simply by adding or removing chiplets within the same package. This plug‑and‑play model allows designers to tailor compute, memory, and I/O resources to specific workload demands—ranging from lightweight edge deployments to high‑performance data‑center configurations—without redesigning a large, complex die. As a result, chiplet architectures can grow or shrink in a granular, cost‑effective manner, matching performance targets and market needs with minimal recurring engineering overhead. However, there is still some trade‑offs between chiplet‑based and monolithic accelerator designs \cite{caichipletmapping}. Notably, the communication overhead introduced by NoP causes chiplet architectures to suffer a performance penalty compared to monolithic chips, which in turn has motivated subsequent research on chiplet performance optimization\cite{shao2019simba,caichipletmapping,cimchiplet,biglittle,sharma2023heterogeneous,dhingra2025atleusacceleratingtransformersedge}.

\subsection{Chiplet-Based CIM Architecture}
\label{sec:previous_work}
Early frameworks and architectures in this space are largely CNN-oriented and emphasize holistic modeling and co-design across the chiplet system. SIAM~\cite{siam} proposes an end-to-end evaluation framework for ACIM-based system. Big-little~\cite{biglittle} introduces a heterogeneous CIM chiplet architecture that mixes large and small ACIM chiplets to improve efficiency for CNN workloads. Recent efforts extend CIM-chiplet co-design toward Transformer-based workloads, typically optimizing at the chiplet/system level via DSE and coarse-grained partitioning/placement to optimize NoP communication. Sharma et al.~\cite{sharma2023heterogeneous} proposed a heterogeneous architecture integrating ReRAM-based ACIM chiplets and streaming multiprocessor chiplets via a network-on-interposer. The placement of chiplets and interconnects are strategically optimized to align with Transformer dataflow requirements. To mitigate the inter-chiplet communication overheads, Jaiswal et al. introduced HALO~\cite{jaiswal2024halo}, a communication-aware 2.5D system designed with a mix of analog CIM and digital floating-point unit chiplets, featuring a greedy algorithm that minimizes NoP movement by mapping model layers to heterogeneous chiplets based on inter-layer traffic. This work, Hemlet, adopts a fully CIM-based chiplet architecture, where different types of chiplets---ACIM and DCIM---are placed on separate chiplets. Building on this hardware architcure, Hemlet further proposes a novel ACIM mapping optimization and a system-level dataflow optimization to improve execution efficiency for ViT workloads.

\section{Hemlet Architecture}
\label{sec:architecture}

\begin{figure*}
    \centering
    \includegraphics[width=\linewidth]{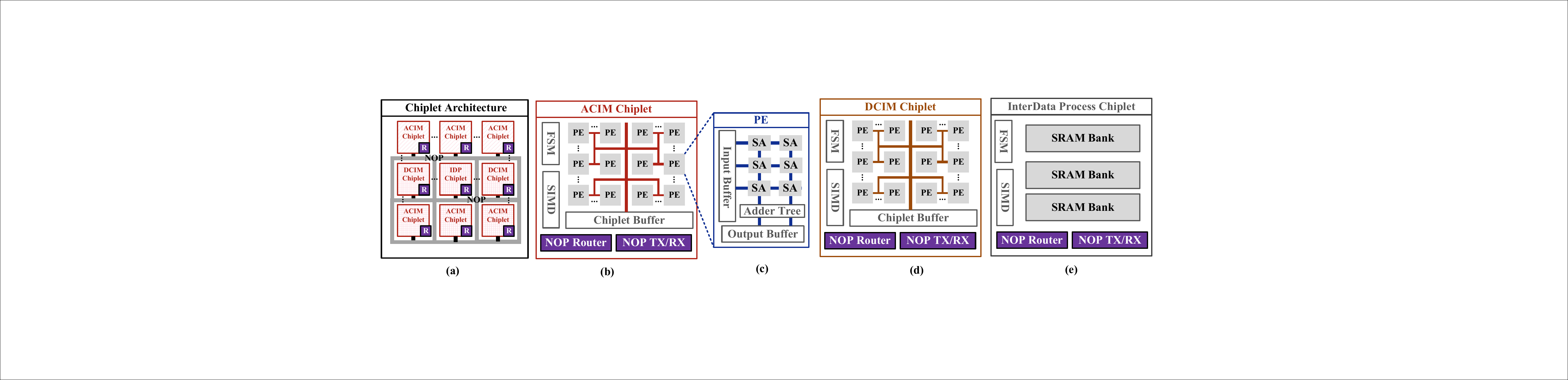}
    \caption{(a) Overall system architecture integrating ACIM, DCIM, and IDP chiplets interconnected through a network-on-package (NoP);
(b) Internal structure of an ACIM chiplet consisting of multiple ACIM PEs, a local buffer, and NoP router and TX/RX;
(c) architecture of a processing engine (PE) with multiple subarrays (SAs), the adder tree, and input/output buffers; (d) Architecture of a DCIM chiplet comprising multiple DCIM PEs. The chiplet also integrates a SIMD unit, a control FSM, a chiplet buffer, and a NoP communication module. (e) Design of an Intermediate Data Process (IDP) chiplet equipped with SRAM banks, a SIMD unit, and NoP communication modules.}
\label{fig:architecture}
\end{figure*}

The detailed Hemlet hardware architecture, illustrated in Fig.~\ref{fig:architecture} (a), comprises three types of specialized chiplets—ACIM, DCIM, and IDP. All chiplets are interconnected through a NoP. To enable seamless inter-chiplet communication, each chiplet integrates a NoP Router along with NoP transmitters (TX) and receivers (RX)—following the design principles of prior works\cite{shao2019simba,siam,biglittle}. In addition, each chiplet integrates a finite state machine (FSM) for control flow, and a single instruction multiple data (SIMD) unit for vector computation (e.g., LayerNorm, GELU, Softmax, accumulation, etc.), and a chiplet buffer for temporary data storage.

\textbf{Analog CIM (ACIM) Chiplet:}
The ACIM chiplet, composed of analog RRAM-based ACIM arrays, functions as the computational engine for static VMMs. A sufficient number of ACIM chiplets are allocated to ensure that the weight matrices of the entire ViT model can be fully stored. Each ACIM chiplet features a two-layer architecture comprising processing engines (PEs) and subarrays (SAs), as illustrated in Fig.~\ref{fig:architecture} (b)-(c).

\textbf{Digital CIM (DCIM) Chiplet:}
DCIM chiplets utilize SRAM-based digital CIM macros as processing engines (PEs), as illustrated in Fig.~\ref{fig:architecture} (d). The DCIM chiplets are shared among the MHA modules from different blocks of the ViT, resulting in lower hardware occupancy compared to the ACIM chiplets. The capacity of the DCIM is tailored to align with the granularity of the operation for dynamic VMMs, which will be discussed in detail in Section~\ref{sec:dataflow}.

\textbf{Intermediate Data Process (IDP) Chiplet:}
The Intermediate Data Process (IDP) chiplet functions as a centralized global buffer and coordination unit within the system. It consists primarily of multiple SRAM banks that temporarily store activations exchanged between layers or chiplets. This design mitigates the limited buffer capacity of individual compute chiplets (e.g., ACIM or DCIM) and reduces dependence on off-chip DRAM. In addition to buffering intermediate data, the IDP chiplet also performs certain SIMD-based auxiliary operations. Unlike the SIMD operations within ACIM or DCIM, the SIMD functionality in the IDP is responsible for handling operators that involve operands spanning multiple chiplets.

\section{Group-Level Parallelism (GLP) Weight Mapping Strategy}
\label{sec:glp-big-section}
\begin{figure*}
    \centering
    \includegraphics[width=\linewidth]{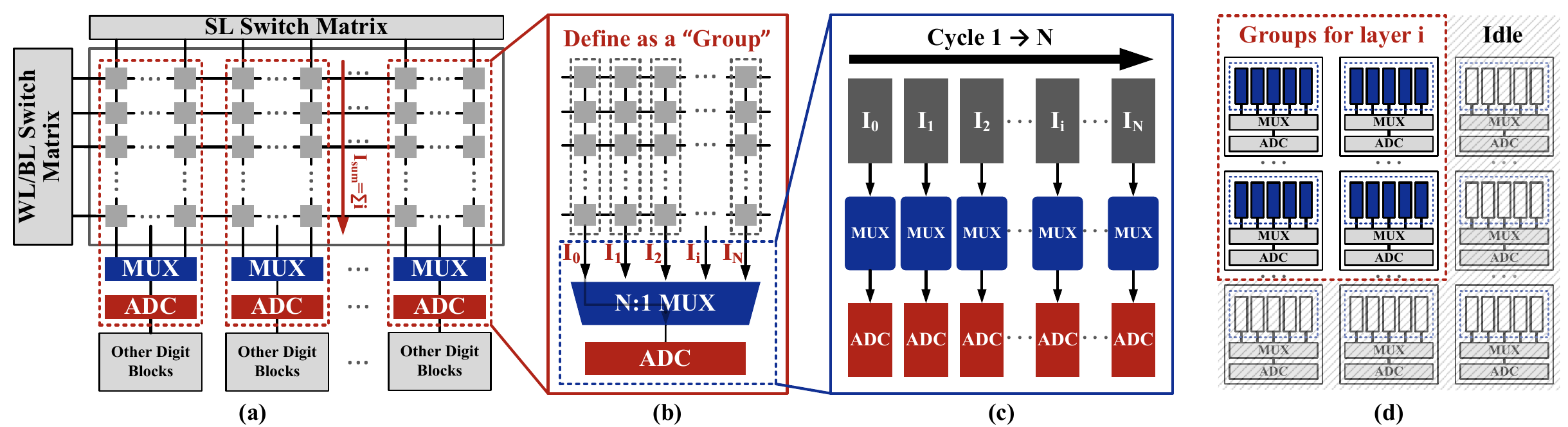}
    \caption{Motivation for group-level parallelism (GLP).
        (a) Typical CIM subarray with MUXs and shared ADCs; (b) A group of columns sharing one ADC and MUX is defined as a “Group”; (c) Time-multiplexed column access severely limits throughput, activating only one column per group per cycle; (d) system-wide ADC under-utilization under the layer-wise mapping method.}
    \label{fig:motivation}
\end{figure*}

\subsection{Group-Level Parallelism: Motivation}

While static VMMs dominate Transformer operations and consequently consume substantial hardware resources in chiplet systems, conventional layer-wise mapping methods—where weights from a single layer are sequentially mapped to the columns of an ACIM array—often do not fully exploit the available ACIM resources during runtime. As described in Section \ref{sec:acim}, although VMMs are performed in the analog domain, ADCs are necessary to convert the results back for further processing. However, in practical implementations, a single ADC is typically shared across multiple memory columns due to physical layout constraints—specifically, ADCs are significantly wider than the pitch of the RRAM columns \cite{shimengreview}. To match the column pitch and reduce ADC area overhead, memory columns are grouped and time-multiplexed via a multiplexer (MUX), allowing only one column per group to access the ADC in each cycle (Fig. \ref{fig:motivation} (b)). While computation occurs concurrently across all columns, the shared ADC digitization process is inherently serialized through multiplexing, significantly constraining the throughput per subarray, shown in Fig. \ref{fig:motivation} (c).

For easy reference, we define the columns that share a single ADC as a \textbf{Group}, shown in Fig.\ref{fig:motivation} (b). Consequently, a single ACIM subarray can be viewed as consisting of multiple such groups. Our key insight is that parallelism in ACIM is achieved at the group level rather than on a column-wise basis. In other words, to fully utilize the ACIM computing resources, analog multiply-and-accumulate (MAC) operations that can be executed concurrently should be mapped to different column groups, while MACs with dependencies can be allocated to the same group since they naturally require access to the ADC at different time steps. Building on this insight, we propose an interleaved mapping strategy called \textbf{Group-Level Parallelism (GLP)}. This approach simultaneously activates significantly more ADC resources across the entire system, thereby mitigating resource under-utilization in ACIM and enhancing its throughput.


\subsection{Cross-layer Interleaved Distribution}
\label{sec:core-idea}
The core idea of GLP is to break away from layer-wise weight mapping by\textbf{ \textit{interleaving weight columns from multiple layers into the same column group}}. This approach aims to increase throughput by allowing more groups to be active during inference of each layer. To illustrate this concept, we first present a toy example based on a layer-by-layer network, without accounting for the properties of Transformers or hardware limitations.

Considering this network has $L$ linear layers, the weight of layer $t$ is denoted by $W^t$. We could collect the $j$-th column vector of $\mathbf{W}^t$ across all $L$ layers to form a column set:
\begin{equation}
\mathcal{C}_{j} = \{ \mathbf{W}^{0}[:,j],\, \mathbf{W}^{1}[:,j],\, \dots,\, \mathbf{W}^{L-1}[:,j] \}.
\end{equation}

\begin{figure}[t]
    \centering
    \includegraphics[width=3.2in]{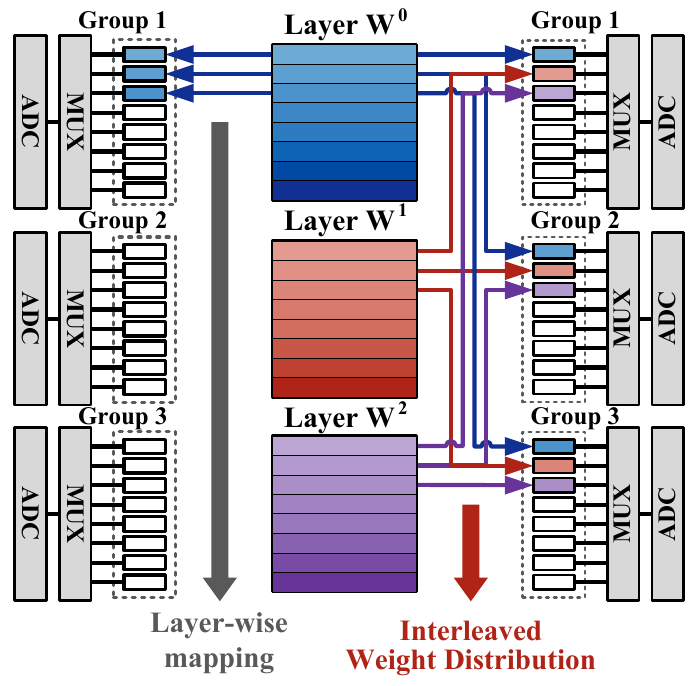}
    \caption{Illustration of the interleaved weight distribution strategy. The layer-wise mapping method (left) statically assigns each group to a specific layer, while the interleaved approach (right) distributes weights from multiple layers across groups to enable concurrent group activation.}
    \label{fig:interleaved-mapping}
\end{figure}

If each $\mathcal{C}_{j}$ is assigned to a column group, as shown in right side of Fig.~\ref{fig:interleaved-mapping}, then for the inference of any layer, all the ADCs will be activated in parallel. To fully leverage this group-level parallelism, certain assumptions must be made. First, the network should exhibit strict layer-by-layer dependencies, with weights of {identical size}. Second, the size of each column group must correspond to the number of layers to maximize column utilization. However, in real-world scenarios, these two assumptions do not naturally hold. In the following section, we will discuss how this method can be adapted for Transformers while taking hardware constraints into account.


\begin{figure}
    \centering
    \includegraphics[width=3.6in]{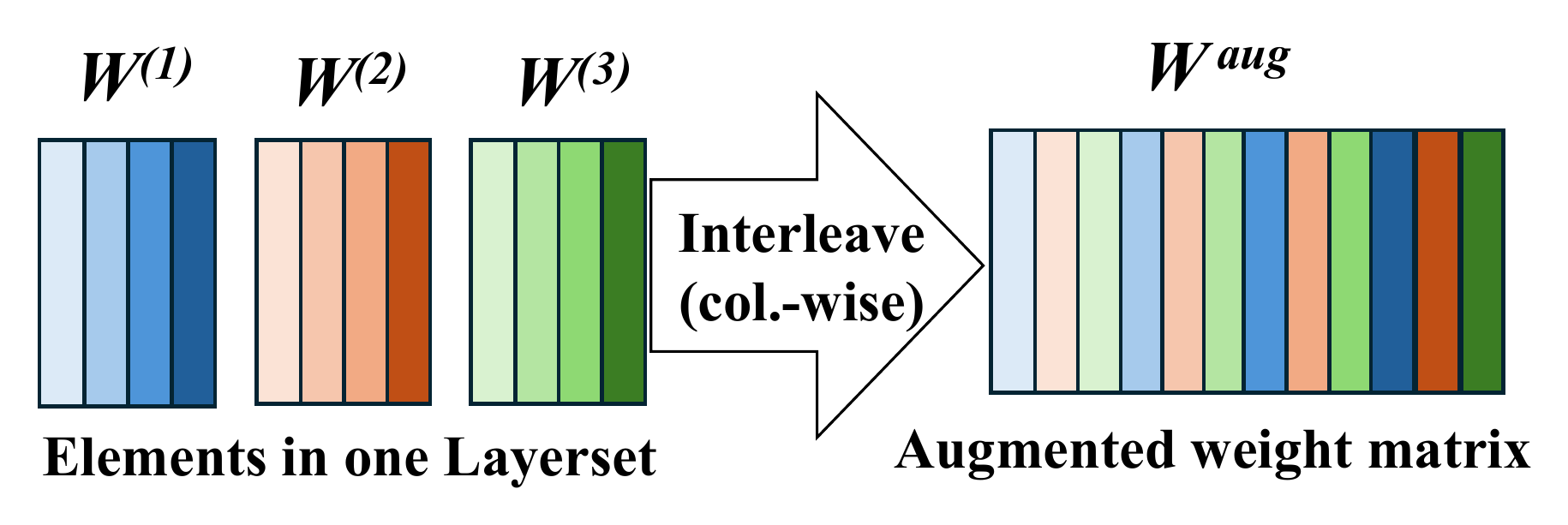}
    \caption{Construction of the augmented weight matrix for each GLP\_LayerSet}
    \label{fig:layerset}
\end{figure}

To integrate GLP into weight mapping, we further introduce a unified intermediate representation (IR) called \textit{GLP\_LayerSet}. Each \textit{GLP\_LayerSet} is composed of several static-weight linear layers from Transformer blocks. During weight mapping, an \textit{augmented weight matrix} is constructed for each \textit{GLP\_LayerSet} by interleaving the column-wise weight matrices of its constituent layers along the column dimension, which is then partitioned and mapped onto CIM arrays. Formally, let $\mathcal{L} = \{\ell_1, \ell_2, \dots, \ell_x\}$ denote a \textit{GLP\_LayerSet} containing $x$ linear layers. Each layer $\ell_i$ has a weight matrix $\mathbf{W}^{(i)} \in \mathbb{R}^{C_\text{in} \times C_\text{out}}$. All layers in $\mathcal{L}$ must satisfy two conditions: (1) their weight matrices have identical dimensions to enable uniform column-wise mapping, and (2) they are not concurrently activated during execution. For a \textit{GLP\_LayerSet}, the weight columns are interleaved to form an augmented weight matrix $W^{\mathrm{aug}}$, as illustrated in Fig.~\ref{fig:layerset}.
\begin{equation}
\mathrm{Interleave}\!\bigl(\mathbf{W}^{(1)},\,\mathbf{W}^{(2)},\dots,\mathbf{W}^{(x)}\bigr)
      \;=\; \mathbf{W}^{\mathrm{aug}}\in
      \mathbb{R}^{C_\text{in}\times xC_\text{out}} .
\end{equation}

\subsection{Deployment Approach}
\label{sec:glp}
As discussed in Section~\ref{sec:core-idea}, two conditions must be satisfied to form a \textit{GLP\_LayerSet}.
Fortunately, the structural properties inherent in Transformer-based architectures facilitate the fulfillment of these conditions. First, Transformer models generally consist of Transformer blocks with the same structure. Second, inside each block, the layers share a regular pattern, which can be easily tailored to identical dimensions. 
As shown in Fig. \ref{fig:transformer}, each Transformer block contains six categories of static weight matrices: the projection layers in the MHA—$\mathbf{W}_Q$, $\mathbf{W}_K$, $\mathbf{W}_V$, and $\mathbf{W}_O$—which all share the same matrix shape of $\mathbb{R}^{d \times d}$, along with two feed-forward network (FFN) layers: $\mathbf{W}_1 \in \mathbb{R}^{d \times D}$ and $\mathbf{W}_2 \in \mathbb{R}^{D \times d}$, where $d$ denotes the embedding dimension and $D$ denotes the FFN hidden dimension.
It can be observed that only $\mathbf{W}_1$ and $\mathbf{W}_2$ differ in size from the other linear layers. In practical implementations, we typically have $D > d$ (for instance, in ViTs, it is common to set $D = 4d$). To align all layers' shapes, this work partitions FFN layers into $k$ sub-layers along the column (for $\mathbf{W}_1$) and row (for $\mathbf{W}_2$) dimensions:
\begin{align}
\mathbf{W}_1 &\rightarrow \{\mathbf{W}1_0, \mathbf{W}1_1, \dots, \mathbf{W}1_{(k-1)}\},\quad \mathbf{W}1_i \in \mathbb{R}^{d \times d} \\ \nonumber
\mathbf{W}_2 &\rightarrow \{\mathbf{W}2_0, \mathbf{W}2_1, \dots, \mathbf{W}2_{(k-1)}\},\quad \mathbf{W}2_i \in \mathbb{R}^{d \times d}
\end{align}
This partition allows all linear layers to be mapped using a unified four-stage \textit{GLP\_LayerSets} construction strategy detailed below. Given a \textit{GLP\_LayerSet} size $M$, which is set to be equal to the group size, and a ViT model with $N$ stacked Transformer blocks:
\begin{enumerate}
\item \textit{\textbf{Stage 1}: FFN Sub-layer Packing.}
The FFN sub-layers are first packed as they dominate the number of parameters. For each sub-layer $\{\mathbf{W}1_i, \mathbf{W}2_i\}$, $i \in \{0,1,\dots,k{-}1\}$, sub-layers with the same $i$ across different blocks are first collected. As a result, there are $k$ collections of $\{\mathbf{W}1_{i}^{0}, \mathbf{W}2_{i}^{0}, \dots, \mathbf{W}1_{i}^{N-1}, \mathbf{W}2_{i}^{N-1}\}$. For each collection, the sub-layers inside it have cross-block dependencies that can be grouped into the same \textit{GLP\_LayerSet}. As a result, each collection is organized into $\lceil 2N/M \rceil$ \textit{GLP\_LayerSets}, and $k \cdot \lceil 2N/M \rceil$ \textit{GLP\_LayerSets} are generated in this stage in total. It is worth noticing that if $2N$ is divisible by $M$, all the \textit{GLP\_LayerSets} are fully filled, and thus Stage 2 can be directly skipped. Otherwise, there will be $k$ \textit{GLP\_LayerSets} that contain empty positions, thus leading to reduced memory efficiency. Therefore, in the following step, MHA layers will be first filled into these slacks.

\item \textit{\textbf{Stage 2: Slack Filling with MHA Layers.}}  
Considering $z = M \cdot \lceil 2N/M \rceil - 2N$ slacks from the previous stage, $z \cdot \lceil k/4 \rceil$ groups of $\{\mathbf{W}_Q, \mathbf{W}_K, \mathbf{W}_V, \mathbf{W}_O\}$ from MHA are filled in. In the ideal case, where $k$ is divisible by $4$, all the layers from MHA are exactly consumed. As $k=4$ is the general case, this filling is commonly effective. For the unusual cases, any remaining MHA layers that are not placed into slack positions during this step are collected as residuals and handled in Stage 4.

\item \textit{\textbf{Stage 3: Processing the MHA layers.}}  
After completing the slack filling in Stage~2, each type of MHA layer in $\{\mathbf{W}_Q, \mathbf{W}_K, \mathbf{W}_V, \mathbf{W}_O\}$ has  
$R$ remaining layers.  
We compute $(p, q) = \text{divmod}(R, M)$.  
If $p \geq 1$, we create $p$ \textit{GLP\_LayerSets} for each of the four MHA types. 
Moreover, if $q > 0$, we handle the $q$ remaining layers of each MHA type as follows: If $3M = 4N$: create three \textit{GLP\_LayerSets} for $\mathbf{W}_Q$, $\mathbf{W}_K$, and $\mathbf{W}_V$, and evenly distribute the $\mathbf{W}_O$ layers among them. Otherwise, all the rest goes to stage 4.

\item \textit{\textbf{Stage 4: Residual Handling.}} The remaining layers that cannot be grouped into any \textit{GLP\_LayerSets} are collected into the \textit{Baseline\_LayerSet} for traditional mapping.
\end{enumerate}

After constructing the \textit{GLP\_LayerSets} from the model, augmented weight matrices are generated as previously described. Given that the size of the \textit{GLP\_LayerSets} is configured to match the group size, every sequential $M$ columns of this matrix can be assigned to a group. As a result, the peripheral ADCs for each \textit{GLP\_LayerSet} can be fully utilized, thereby enhancing parallelism and increasing throughput.

\section{System Dataflow Optimization}
\label{sec:dataflow}
As introduced in Section~\ref{sec:architecture}, Hemlet leverages heterogeneous CIM chiplets—ACIM for static VMMs and DCIM for runtime-generated dynamic VMMs. While this heterogeneous organization offers architectural flexibility, it also amplifies inter-chiplet data movement.

\begin{figure*}
    \centering
    \includegraphics[width=\linewidth]{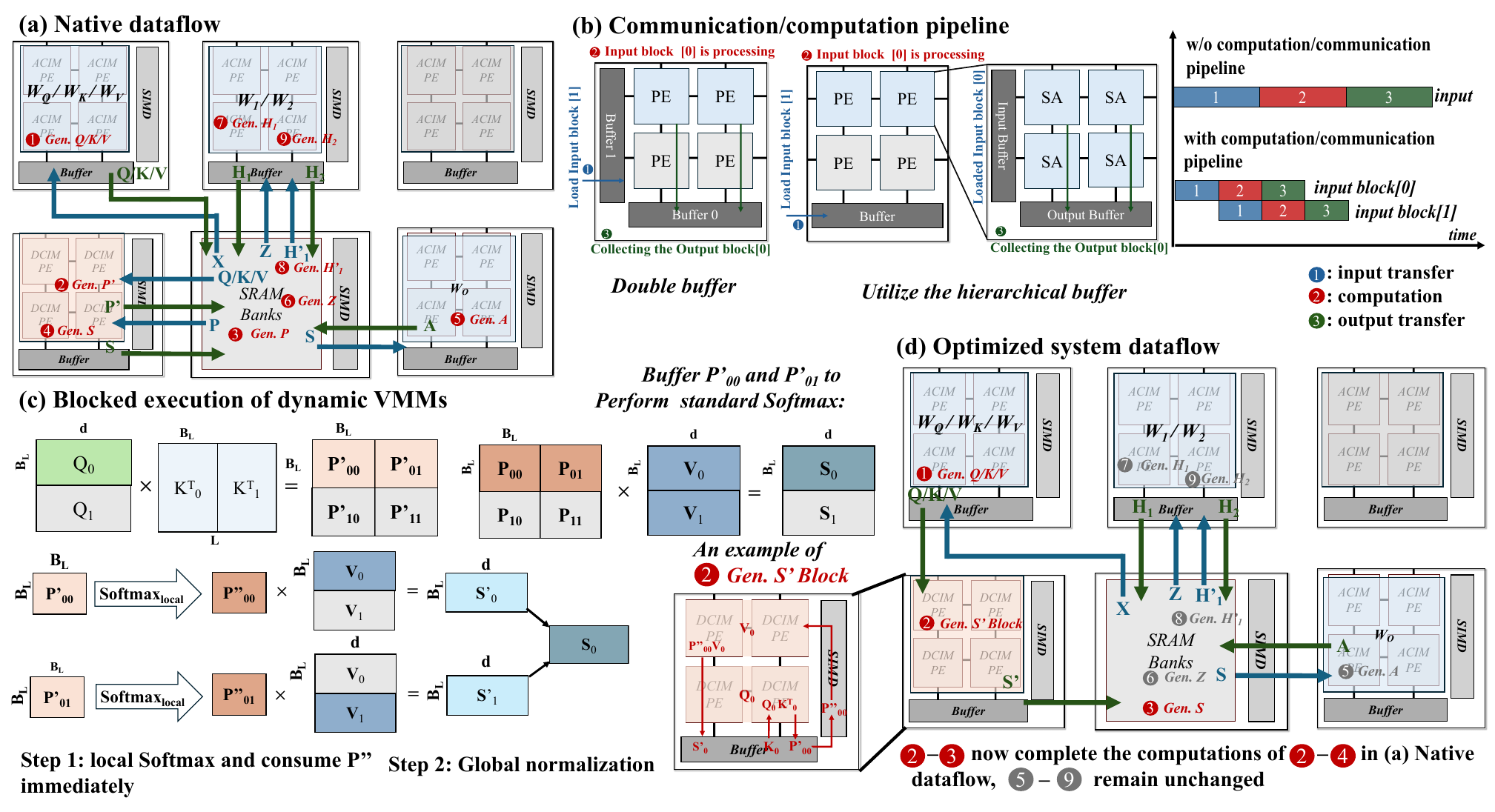}
    \caption{Dataflow Optimization. 
(a) Native dataflow for a Transformer block (basic layer-by-layer execution on heterogeneous chiplets); (b) Communication/computation pipeline (utilize the hierarchical buffers to overlap communication latency); 
(c) Blocked dynamic VMMs on DCIM chiplets (the standard Softmax is split into two stages so each intermediate $\mathbf{P}'$ block can be consumed immediately, followed by a final normalization across blocks); 
(d) Optimized system dataflow (reduced inter-chiplet communication and improved utilization).}
    \label{fig:dataflow}
\end{figure*}


A basic dataflow for Transformers in the heterogeneous chiplet system can be executed in a layer-by-layer manner. Fig. \ref{fig:dataflow} (a) provides a simplified example to illustrate this native dataflow for Transformer blocks shown in Fig. \ref{fig:transformer}. For each Transformer block, \bcirc{1} the hidden states $\mathbf{X} \in \mathbb{R}^{L \times d}$ are loaded from the global buffer in IDP to the corresponding ACIM chiplets for $\mathbf{Q}/\mathbf{K}/\mathbf{V}$ projection, where $L$ denotes the sequence length and $d$ denotes the embedding dimension. These results are then transferred back to the global buffer in IDP chiplet, where it awaits the next processing. Subsequently, \bcirc{2} the $\mathbf{Q}/\mathbf{K}/\mathbf{V}$ are sent to the DCIM chiplets to generate $\mathbf{P'} = \mathbf{Q} \mathbf{K}^T$, which is then \bcirc{3} returned to the IDP for the Softmax operation to obtain  $\mathbf{P} = Softmax(\mathbf{P'})$. After that, \bcirc{4} the $\mathbf{P}$ matrix will be sent back to DCIM for $\mathbf{S}=\mathbf{P}\mathbf{V}$ and the result $\mathbf{S}$ is buffered again in the IDP. It is then \bcirc{5} sent to the ACIM chiplets corresponding to $\mathbf{W}_O$ to perform the output projection and generate $\mathbf{A}$. \bcirc{6} The resulting $\mathbf{A}$ is written back to IDP, where residual addition and LayerNorm are applied to generate $\mathbf{Z}$ before passing it to FFN. After that, \bcirc{7} the $\mathbf{Z}$ is then sent to the ACIM chiplets for $\mathbf{W}_1$ to obtain the result $\mathbf{H}_1$. \bcirc{8} The outputs $\mathbf{H}_1$ are collected back to the IDP, where the GELU activation is applied to obtain $\mathbf{H'}_1$. This \bcirc{9} activated result $\mathbf{H'}_1$ is subsequently forwarded to the ACIM chiplets for $\mathbf{W}_2$, producing $\mathbf{H}_2$. Finally, $\mathbf{H}_2$ is written back to the IDP, where residual connection and LayerNorm will be performed again, yielding the hidden states for the next Transformer block. This dataflow, while simple, incurs the data movement across chiplets since the static VMMs, dynamic VMMs and global buffer are located on different chiplets in Hemlet.

For the dataflow employed by Hemlet, pipeline across blocks of inputs is employed for static VMMs on the ACIM chiplets. Specifically, input hidden states $\mathbf{X} \in \mathbb{R}^{L \times d}$ are cut into $L / B_L$ temporal blocks along the sequence length ($L$). The process and transfer of these blocks are pipelined utilizing the double buffering scheme favored by the hierarchical buffering system of ACIM chiplets, as illustrated in Fig. \ref{fig:dataflow} (b). Rather than doubling the buffer size, we utilize the hierarchical buffering system of the CIM chiplet to enable pipelining. Specifically, for inter-chiplet communication, the chiplet buffer serves as both the source and the destination of data transfer. For the processing, the data is subsequently transferred to a local buffer surrounding the CIM macros for computation. This approach allows the chiplet buffer to be overwritten with the next chunk of data, optimizing resource utilization.

While this communication/computation pipeline helps hide some of the communication latency, it does not reduce actual data movement. Moreover, it is not easy to directly apply the same scheme to DCIM due to data dependency in attention. To address this, we introduce a design optimization for DCIM chiplets as shown in Fig. \ref{fig:dataflow} (c). Specifically, the DCIM chiplet is intentionally designed with a larger buffer capacity, sufficient to hold the entire $\mathbf{Q}/\mathbf{K}/\mathbf{V}$ data required by the heads processed on this DCIM chiplet. This \bcirc{1} allows the $\mathbf{Q}/\mathbf{K}/\mathbf{V}$ outputs generated by the ACIM chiplets to be transmitted directly to the DCIM chiplets, thereby eliminating the need for intermediate storage in the IDP chiplet. This design is viable because DCIM chiplets are shared across all Transformer blocks and thus could bear the relatively larger buffer overhead.

For the dynamic VMMs operation, the input matrices $\mathbf{Q}$,  $\mathbf{K}$ and $\mathbf{V}$ are partitioned into smaller blocks along the sequence length for blocked processing considering the DCIM capacity limit, adopting the same block size $B_L$ as used in the previous stage to match the processing granularity. For the in-memory VMMs, blocks of $\mathbf{Q}$ and $\mathbf{V}$ will be written into the DCIM array as weights for $\mathbf{Q}\mathbf{K}^T$ and $\mathbf{P}\mathbf{V}$ operations separately. To reduce the forward and backward transfers of the intermediate matrix $\mathbf{P}'$ between the DCIM chiplet and the IDP, we adopt a placement strategy that prioritizes mapping the PEs responsible for both $\mathbf{Q}\mathbf{K}^T$ and $\mathbf{P}\mathbf{V}$ of the same attention head to the same DCIM chiplet. While the process is blocked, fully executing the Softmax operation between $\mathbf{Q}\mathbf{K}^T$ and $\mathbf{P}\mathbf{V}$ on local DCIM chiplets requires a much larger chiplet buffer to store the entire intermediate matrix $\mathbf{P}'$ of size $B_L \times L$. To reduce this buffer overhead, we adopt an optimization inspired by FlashAttention~\cite{flashattention}, a local $\mathrm{Softmax}_{\text{local}}$ is applied to each block result of $\mathbf{P'} = \mathbf{Q}\mathbf{K}^T$, yielding $\mathbf{P''} = \mathrm{Softmax}_{\text{local}}(\mathbf{P'})$. The corresponding block output is then obtained as \bcirc{2} $\mathbf{S'} = \mathbf{P''}\mathbf{V}$ in Fig. \ref{fig:dataflow} (c). After collecting all $\mathbf{S'}$ blocks along the standard Softmax dimension, a \bcirc{3} global normalization is performed to produce the  $\mathbf{S}$ block. In this way, each $\mathbf{P}'$ block can be consumed immediately, avoiding much larger on-chip buffer of the DCIM chiplet or the forward and backward transfers between DCIM chiplets and IDP chiplet.

\section{Evaluation}
\label{sec:evaluation}


\subsection{Experimental Methodology}
\textbf{Simulator and Modeling Framework:}
We develop a custom cycle-accurate simulator tailored for heterogeneous CIM-based chiplet systems. The simulation flow integrates hardware-level modeling, dataflow-level scheduling, and event-driven instruction execution across computations on chiplets and inter-chiplet communication. For ACIM chiplets, we adopt NeuroSim~\cite{NeuroSimv1.3} to model RRAM-based subarrays and other digital blocks at a 22 nm technology node. DCIM chiplets and the IDP are also extended from NeuroSim and assumed at the 7 nm technology node, with DCIM array performance from the AutoDCIM design~\cite{autodcim} and scaled to 7 nm using DeepScale~\cite{sarangi2021deepscaletool}. To estimate inter-chiplet communication cost, we extend BookSim~\cite{jiang2013detailed} for latency estimation under a mesh topology. The corresponding energy consumption is estimated based on the advanced packaging parameters provided in UCIe~\cite{sharma2022ucie}. 
To determine the hardware configuration of ACIM, we utilize MICSim~\cite{micsim} to evaluate ImageNet-1K\cite{imagenet} classification performance under the \texttt{ViT-B/16} model. We perform quantization-aware training (QAT) with I-ViT\cite{li2023ivit}, achieving $80\%$ accuracy. Our results indicate that with an ADC precision of 9 bits and 2-bit RRAM with an $R_{on}/R_{off}$ ratio of 150\cite{rram150}, no accuracy degradation is observed compared to our software baseline.

\textbf{System Configuration:}
The system is configured to ensure all pretrained weights are fully mapped into ACIM chiplets before inference, consistent with prior works~\cite{retransformer,x-former,shafiee2016isaac,chi2016prime,siam}. An ACIM chiplet consists of multiple ACIM PEs. Each ACIM PE features a fixed configuration of $60$ ACIM subarrays ($128\times128$). The group size for the ACIM subarray is set to 8. The buffer size of each ACIM chiplet is 64 KB. Meanwhile, the DCIM chiplet comprises multiple DCIM PEs, with each DCIM PE integrating 4 DCIM subarrays ($64\times64$). The buffer size of each DCIM chiplet is 512 KB. 
To account for the varying capacities of chiplets, we assume that the number of ACIM PEs for the ACIM chiplet and the number of DCIM PEs for the DCIM chiplet may vary. To analyze the impact of chiplet size and communication bandwidth, we evaluate configurations of A18D9 (each ACIM chiplet with {18 ACIM PEs} and each DCIM chiplet with {9 DCIM PEs}), A32D16, and A50D25 under a range of NoP bandwidths from 8 GB/s to 32 GB/s. The NoP communication network is configured in a 2D mesh topology, which follows the design in Simba\cite{shao2019simba}.

\begin{table}[t]
\caption{Vision Transformer Model Configuration}
\label{tab:vit-models}
\centering
\begin{tabular}{lcccccc}
\toprule
\textbf{Model} & \textbf{Dim.} & \textbf{Heads} & \textbf{Blocks} & \textbf{Patch Size} & \textbf{Seq. Len.} \\
\midrule
ViT-S/16 & 384 & 6 & 12 & $16 \times 16$ & 197 \\
ViT-B/16 & 768 & 12 & 12 & $16 \times 16$ & 197 \\
ViT-L/16 & 1024 & 16 & 24 & $16 \times 16$ & 197 \\
\bottomrule
\end{tabular}
\end{table}


\begin{figure*}
    \centering
    \includegraphics[width=\linewidth]{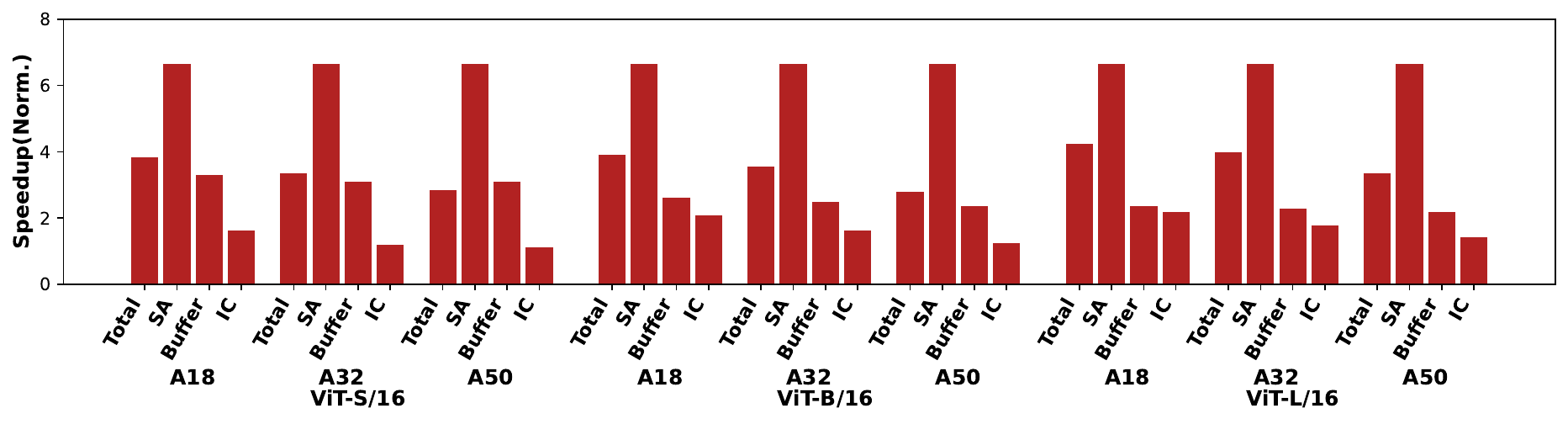}
    \caption{The GLP speedup breakdown of ACIM chiplet}
    \label{fig:acim_speedup}
\end{figure*}


\textbf{Workloads:} 
We evaluate our system using standard ViT models~\cite{vit}, covering three representative scales: \texttt{ViT-S/16}, \texttt{ViT-B/16}, and \texttt{ViT-L/16}. These models span a wide range of parameter counts and computational requirements, allowing us to assess the scalability and efficiency of our proposed architecture and GLP strategy. All models are evaluated under the image resolution of $224 \times 224$, using $16\times16$ patching, and are fully quantized to 8-bit with I-ViT~\cite{li2023ivit} on the ImageNet-1K\cite{imagenet}. The detailed model parameters, including embedding dimension, number of Transformer blocks, and the Sequence length, are summarized in Table~\ref{tab:vit-models}. 


\subsection{Impact of GLP on ACIM Chiplet Execution}
\label{sec:acim_result}

To validate the effectiveness of GLP on the internal execution behavior of ACIM chiplets, we perform a fine-grained profiling of static VMMs under both baseline (layer-wise weight mapping) and GLP mapping strategies. Peak performance is evaluated, where the input activations are assumed fully loaded into the destination ACIM chiplet’s buffer, isolating the core execution latency and eliminating external communication interference. 
Fig. \ref{fig:acim_speedup} shows the normalized speedup of GLP over the baseline across three ViT models and hardware configurations. A speedup breakdown is also presented for (i) ACIM subarrays (SA), (ii) hierarchical buffer accesses (Buffer), and (iii) multi-level interconnect (IC).

Overall, GLP achieves consistent performance improvements across all settings, primarily driven by improvements in subarray execution. 
This outcome aligns with the design objective of GLP, which interleaves weights across memory groups to improve the SA level parallelism. 
In addition, both buffer and IC exhibit performance improvements under GLP mapping. This benefit stems from the weight column-wise interleaving strategy in GLP, which activates more groups and thereby enables concurrent operation of subarrays, PEs, and even chiplets. As a result, the output data are more evenly distributed across each hierarchy, which in turn reduces buffer pressure and shortens communication latency.

Above results demonstrate that GLP significantly improves execution efficiency within ACIM chiplets. It not only eliminates the bottleneck of time-multiplexed ADCs in ACIM, but also reduces buffer and interconnect pressure via finer-grained output partitioning. These benefits are consistent across various ViT models and chiplet configurations, confirming the scalability and generality of GLP in ACIM execution.

\begin{figure*}
    \centering
    \includegraphics[width=\linewidth]{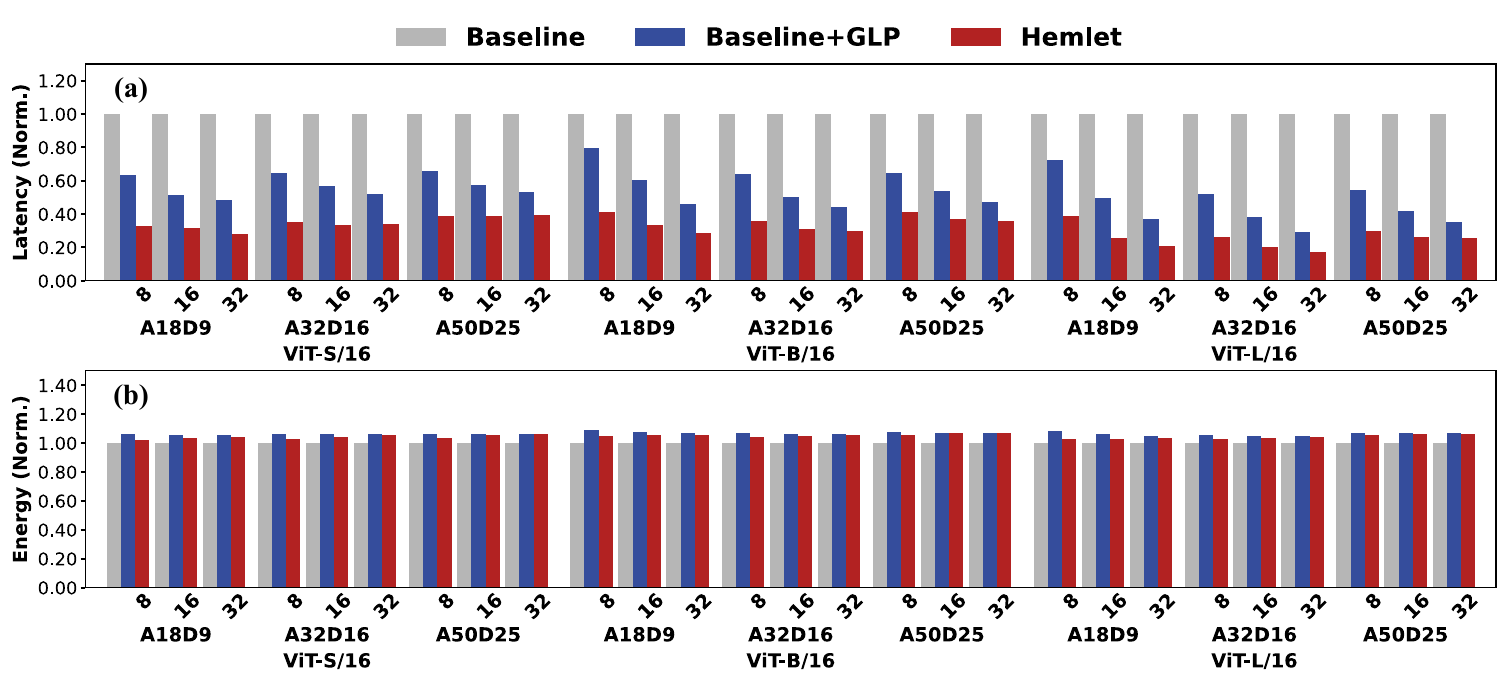}
    \caption{Normalized system latency and energy under different chiplet configurations and NoP bandwidths (8/16/32 GB/s).}
    \label{fig:system_performance}
\end{figure*}

\subsection{System-level Performance Analysis}
 
\textbf{Ablation Study:} 
To evaluate the effects of the proposed GLP mapping and system dataflow optimization on the system performance, we perform an ablation study across different hardware settings. The analysis considers three execution strategies: \textbf{Baseline}, which adopts the conventional layer-wise mapping with the native system dataflow described in Section~\ref{sec:dataflow}; \textbf{Baseline + GLP}, which further applies the proposed GLP mapping; and \textbf{Hemlet}, which incorporates both GLP and system-level dataflow optimization. 

Fig.~\ref{fig:system_performance}(a) shows that GLP reduces latency in most settings by improving ACIM throughput, with the maximum system speedup reaching $2.53\times$. The effectiveness of this improvement is sensitive on both NoP bandwidth and chiplet size. Under lower bandwidth NoP configurations, inter-chiplet communication occupies a larger fraction of the overall execution time, which limits the achievable benefit from ACIM-side acceleration. As the NoP bandwidth increases, the communication bottleneck is gradually alleviated, allowing the throughput advantages of GLP-enabled ACIM execution to translate more effectively into system speedup. The system-level dataflow optimization further improves performance by alleviating inter-chiplet communication, providing greater benefits especially under conditions of limited bandwidth and smaller chiplet sizes. Overall, the proposed Hemlet architecture, which integrates GLP and system-level dataflow optimization, ensures consistent acceleration across all tested configurations, delivering system-level speedups ranging from $2.41\times$ to $5.74\times$.

As shown in Fig.\ref{fig:system_performance} (b), we can observe that GLP mapping introduces extra energy overhead in all settings. This arises because, under GLP, the weights of each layer are interleaved across more groups, leaving each ACIM subarray with fewer weight columns. As a result, each input is reused less within one subarray and must be loaded into more subarrays, thereby increasing data movement and overall energy consumption. After applying the system-level dataflow optimization, the overall energy consumption can be reduced. This is because the optimization decreases inter-chiplet communication, thereby lowering the energy cost of data transfers. As a result, the system-level dataflow optimization can partially compensate for the additional energy overhead introduced by GLP, with an average overhead of $4.62\%$.

\textbf{Comparison with Existing Work:} 
In Table \ref{tab:comparison}, we compare our design with two state-of-the-art ViT accelerators. The reported results for Hemlet correspond to the configuration using the A32D16 chiplet setup with 32 GB/s NoP bandwidth, evaluated on the \texttt{ViT-L/16} model for end-to-end inference, where each processed element in non-VMM operations is counted as one operation. The work presented in \cite{dac24dcim} utilizes DCIM arrays as the computation cores for both static VMMs and dynamic VMMs. While the reload capability of the DCIM computation units allows for a monolithic design with limited on-chip resources, this approach results in relatively low throughput due to the necessary weight reloads. On the other hand, H3DAtten \cite{li2023h3datten} shares the same design principles as Hemlet, utilizing heterogeneous computing units and employing ACIM arrays to store all weights on-chip. To address area constraints, they employ 3D integration of the eNVM arrays with the other digital circuits. This 3D integration mitigates technology lock-in, as the top and bottom tiers can be fabricated using different process nodes. Hemlet achieves superior throughput compared to H3DAtten, which can be mainly attributed to two key reasons. First, H3DAtten employs a lower precision ADC, necessitating a reduction in row parallelism in the CIM computation to maintain accuracy. Second, Hemlet enhances throughput further by implementing GLP and optimizing dataflow at the system-level. H3DAtten demonstrates better energy efficiency than Hemlet, which can also be attributed to two primary factors. First, the 3D integration in H3DAtten allows for a simpler NoC and incurs lower communication overhead compared to the NoP adopted by Hemlet. Second, H3DAtten locates its ADCs on the bottom tier and fabricates them using a more advanced technology node. In contrast, Hemlet integrates the ADCs on the ACIM chiplet, sharing the same technology node as the eNVM arrays. While this 3D integration in H3DAtten enhances energy efficiency, it also introduces greater design complexity and reduces flexibility due to the tight coupling between the top-tier array and the bottom-tier ADC.

\begin{table}[t]
\centering
\caption{Comparison of recent CIM accelerator designs.}
\label{tab:comparison}
\resizebox{0.7\linewidth}{!}{   
\begin{threeparttable}
\scriptsize
\begin{tabular}{ccccc}
\toprule
\textbf{Work}   & \textbf{DAC'24\cite{dac24dcim}} & \textbf{H3DAtten\cite{li2023h3datten}}   & \textbf{Hemlet} \\
\midrule
Technology (nm)   & 22 & 16 \& 40  & 7 \& 22 \\
Frequency (MHz)    & 500 & 185  & 500\\
\multirow{2}{*}{Hardware} 
             & DCIM & ACIM/DCIM   & \textbf{ACIM/DCIM} \\
              & Monolithic  & 3D integration  & \textbf{2.5D Chiplet} \\
Precision    & INT8 & INT8 & INT8 \\

ADC Precision & N/A & 4-bit & 9-bit\\

Throughput       & \multirow{2}{*}{0.828} & \multirow{2}{*}{1.61}  & \multirow{2}{*}{\textbf{9.56}}    \\
(TOPS)  \\
Energy Efficiency      & \multirow{2}{*}{N/A} & \multirow{2}{*}{7.1} & \multirow{2}{*}{\textbf{4.98}}  \\
(TOPS/W) \\
\bottomrule
\end{tabular}
\end{threeparttable}
} 
\end{table}

\section{Conclusion}
\label{sec:conclusion}
In this work, we present Hemlet, a heterogeneous compute-in-memory chiplet architecture tailored for efficient and scalable ViT inference. Recognizing the inherent limitations of traditional monolithic CIM implementations, our architecture strategically integrates ACIM, DCIM, and IDP chiplets interconnected via a NoP, enabling modular scaling of both compute and memory resources. To address the critical throughput bottleneck found in ACIM chiplet systems, we introduce the group-level parallelism (GLP) mapping strategy. By interleaving weights across multiple column groups, GLP effectively resolves micro-architectural ADC serialization bottlenecks and significantly enhances ACIM resource utilization. Further, we propose a system-level dataflow optimization to reduce the inter-chiplet overhead and improve utilization. 

Our evaluations demonstrate that Hemlet effectively improves the efficiency and scalability of ViT inference. The proposed GLP mapping further enhances ACIM throughput at the cost of additional energy overhead, while system-level dataflow provides consistent performance improvement. Moreover, Hemlet achieves competitive performance compared to state-of-the-art accelerators, offering flexibility and scalability through its heterogeneous design. Overall, Hemlet demonstrates a practical path toward high-performance ViT acceleration.


\bibliographystyle{ACM-Reference-Format}
\bibliography{ref}


\end{document}